\documentclass[letter,twocolumn]{jpsj2}
\usepackage{epsfig}
\title{Stripe State in the Lowest Landau Level}
\author{Naokazu {\sc Shibata} and Daijiro {\sc Yoshioka}}
\inst{
Department of Basic Science, University of Tokyo,
Komaba 3-8-1 Meguro-ku, Tokyo 153-8902, Japan
}
\recdate{October 28, 2003}

\abst{The stripe state in the lowest Landau level is studied
by the density matrix renormalization group (DMRG) method.
The ground state energy and 
pair correlation functions are systematically calculated 
for various pseudopotentials in the lowest Landau level. 
We show that the stripe state
in the lowest Landau level is realized only in  
a system whose width perpendicular to the 
two-dimensional electron layer is
smaller than the order of magnetic length.} 

\kword{fractional quantum Hall effect, stripe,   
ground state, phase diagram, two dimension, density matrix,
renormalization group}

\def\runauthor{Naokazu {\sc Shibata} and Daijiro {\sc Yoshioka}}

\begin{document}

\maketitle

\section{Introduction}
The two-dimensional electron systems in a high magnetic field 
exhibit many interesting properties.
The transport measurements on quantum Hall systems
show the presence of zero-resistance states
at various fractional fillings $\nu=n/(2n\pm 1)$\cite{exp3,FQHE},
which is due to the formation of incompressible
liquid states\cite{Lagh}, while 
finite conductance is observed at $\nu=1/2=\lim_{n\rightarrow \infty}
n/(2n\pm 1)$ 
even in the limit of low temperature\cite{exp2}, 
which suggests the formation of compressible liquid state
predicted in the composite fermion theory\cite{jain}.
Transitions to insulating states have also been observed
at low fillings around $\nu\sim 1/5$\cite{exp5_1,exp5_2,exp5n},
and they are thought to be caused by the formation of Wigner 
crystal, which is pinned by impurity potentials.

Analytical and numerical studies\cite{yoshi,jain,Halp,Onod,Lam,Kun}
have confirmed the existence of these ground states 
in the lowest Landau level, but there still remain questions 
on the ground state between the fractional quantum Hall states
at $\nu=n/(2n\pm 1)$.
In particular, recent DMRG calculations show the presence of 
a weak stripe state in the lowest Landau level
around $\nu=0.42$, 0.37 and below 0.32,\cite{Shib2}
whose correlation functions are different from
those expected in the composite fermion theory\cite{CFSt},
while recent experiments on ultra high-mobility wide quantum wells
did not yield any evidence of the stripe formation in the lowest 
Landau level.\cite{WQW}

In the present paper, we show that the stripe state
in the lowest Landau level is realized only in  
a system of narrow quantum wells whose width is 
smaller than the order of magnetic length.
There is a phase transition to a liquid state as the width 
increases, which is caused by a decrease in short-range 
pseudopotentials.
This may be a reason why the stripe state in the lowest Landau 
level has not been observed in wide quantum wells.

We calculate the ground-state wavefunction by employing the
density-matrix renormalization-group (DMRG) method,\cite{DMRG}
which is applied to two-dimensional 
systems in a magnetic field.\cite{Shib,DY,Shib3} 
This method enables us to obtain the essentially exact 
ground state of large size systems extending the limitation of
the exact diagonalization method with controlled accuracy.

\section{Model and Method}
We use the Hamiltonian of two-dimensional electrons in 
a high perpendicular magnetic field. 
The kinetic energy of the electrons is quenched 
by the magnetic field and we can omit this energy.
Thus the Hamiltonian contains only the Coulomb interaction
\begin{equation}
H=\sum_{i<j} \sum_{\bf q} e^{-q^2/2} V(q) \ 
e^{i{\bf q} \cdot ({\bf R}_i-{\bf R}_j)} 
\label{Coulomb}
\end{equation}
where ${\bf R}_i$ is the guiding center coordinate of the $i$th
electron, which satisfies the commutation relation,
$[{R}_{j}^x,{R}_{k}^y]=i\ell^2\delta_{jk}$.
$V(q) =2\pi e^2/\varepsilon q$ is the Fourier transform of the
Coulomb interaction.
The magnetic length $\ell$ is set to be 1 and
we use $e^2/\varepsilon \ell$ as units of energy.
We assume that the magnetic field is strong enough to 
polarize the spins and suppress the Landau level mixing.

In the present DMRG calculations, we divide the system into 
unit cells $L_x\times L_y$ with the periodic boundary 
conditions for both $x$- and $y$-directions.
We calculate the ground-state energy and wavefunction for systems
with up to 24 electrons in the unit cell with various aspect 
ratios $L_x/L_y$, and 
obtain the pair correlation function $g({\bf r})$
defined by
\begin{equation}
g({\bf r}) \equiv \frac{L_x L_y}{N_e(N_e-1)}\langle 
\Psi | \sum_{i\neq j} \delta({\bf r}+{\bf R}_i-{\bf R}_j)|\Psi
\rangle
\end{equation}
from the ground state wavefunction $|\Psi\rangle$.
The correlation functions in the 
stripe state are calculated in the 
unit cell which has the minimum energy with 
respect to $L_x/L_y$.
These correlation functions are expected to have the 
correct period of the stripes realized in the 
thermodynamic limit.

The accuracy of the results is controlled by 
the density matrix used in the calculation. The truncation 
error in the norm of the wavefunction obtained in the 
present calculation is typically $10^{-4}$ 
with keeping 200 eigenstates of the density matrix. 

\section{Stripe Ground State at $\nu=3/8$}
The stripe state in the lowest Landau level 
has already been obtained in the previous DMRG study,\cite{Shib2}
which shows the enhancement of the stripe 
correlation between the incompressible liquid states
around $\nu\sim 0.42$, $0.37$ and between $\nu\sim 0.32$ and 
$0.15$.
Here, we present detailed results at $\nu=3/8$.

\begin{figure}[t]
\epsfxsize=80mm \epsffile{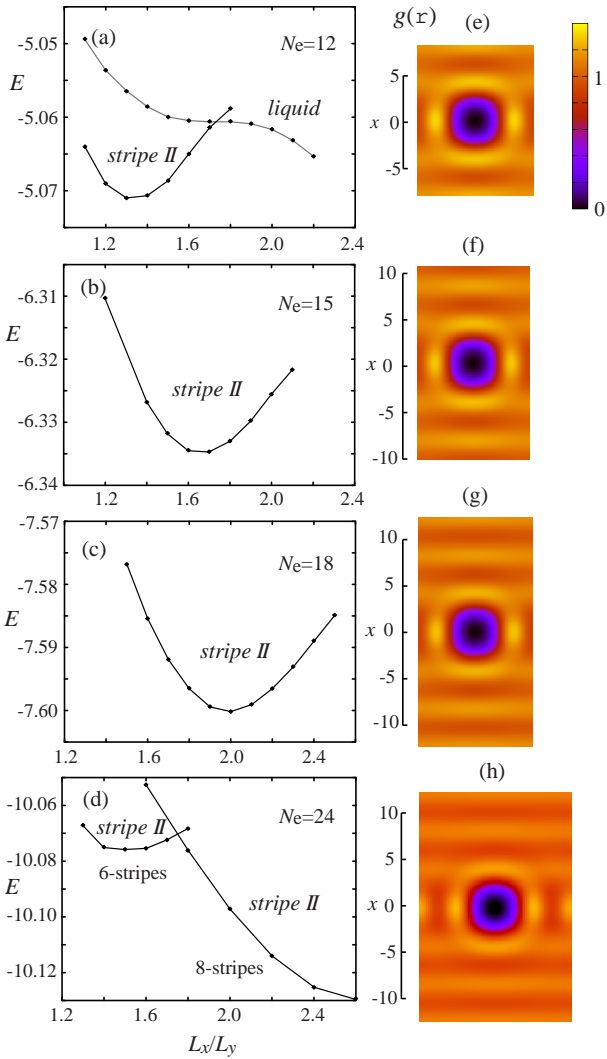}
\caption{
The ground state energy and the pair correlation 
functions for various sizes of systems at
$\nu=3/8$ in the lowest Landau level.       
$N_e$ is the number of electrons in the unit cell
of $L_x\times L_y$.
{\it stripe-II} is the stripe state in the lowest Landau level.
}
\end{figure}

Figure~1 shows the ground-state 
energy and the pair correlation 
functions at $\nu=3/8$ obtained for the systems of 
$N_e=12, 15, 18$ and $24$, where $N_e$ is the number 
of electrons in the unit cell.
We find that the stripe state has the lowest energy
independent of the number of electrons.
There is a liquid ground state in the 
system of 12 electrons at the aspect ratio 
larger then 1.8, but the energy is higher
than that of the stripe state around $L_x/L_y\sim 1.3$
and we think this liquid state is not realized
in large systems.

The energy of the stripe state in Fig.~1
clearly depends on the aspect ratio.
This clear dependence shows
the existence of an optimum stripe structure
because the change in the aspect ratio 
modifies the period of the stripes.
Indeed, the pair correlation functions 
calculated at the energy minimum have almost the same
stripes structure as shown in Figs.~1 (e) $-$ 1 (h).
Since the size dependence is small, 
we expect the existence of the stripes even in the 
thermodynamic limit. In order to examine the size dependence 
in more detail, we plot the correlation functions along the 
$x$-axis for various sizes of systems.
The results for $N_e=12$, 15, 18 and 24
presented in Fig.~2 (a) show that the size dependence 
on the amplitude and the period of the stripes is very small,
which strongly suggests that almost the same 
correlation function is realized in the thermodynamic limit.
We note that the period of the stripes is about $4\ell$, which 
is much smaller than $10\ell$ expected in the composite 
fermion theory.\cite{CFSt}

\begin{figure}[t]
\epsfxsize=70mm \epsffile{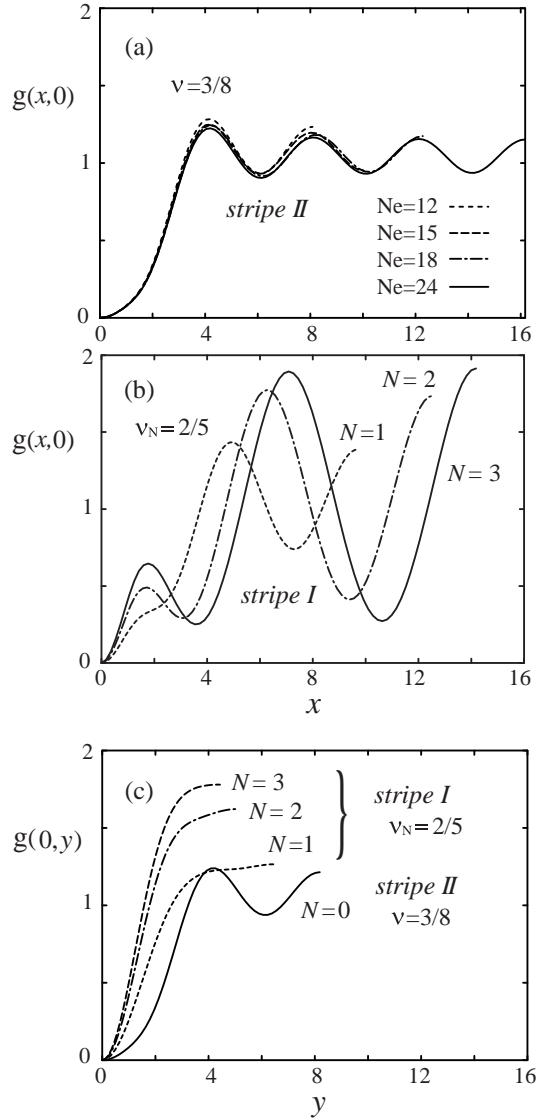}
\caption{(a) The size dependence of the pair correlation
functions at $\nu=3/8$. (b) The pair correlation
functions in the stripe state in higher Landau levels.
$\nu_N$ is the filling of partially filled 
$N$th Landau level.
(c) The correlation functions along the $y$-axis.
{\it stripe-I} is the stripe state near half filling 
in higher Landau levels.
}
\end{figure}

For comparison with the stripe state in 
higher Landau levels, we present in Fig.~2 (b)
the correlation functions in higher Landau levels.
We find that the amplitude of the stripes in the 
lowest Landau level is relatively small
and the short range correlations around 
$r\sim 2\ell$ are different. 
Furthermore, the stripe state in the lowest Landau level has 
clear oscillations along the stripes as shown in 
Fig.~2 (c).
Since the correlation functions are 
qualitatively different, we call the stripe state 
in the lowest Landau level, type-II stripe state.
All the stripe states found in the lowest Landau level 
are characterized by the type-II stripe state.\cite{Shib2} 

\begin{figure}[t]
\epsfxsize=80mm \epsffile{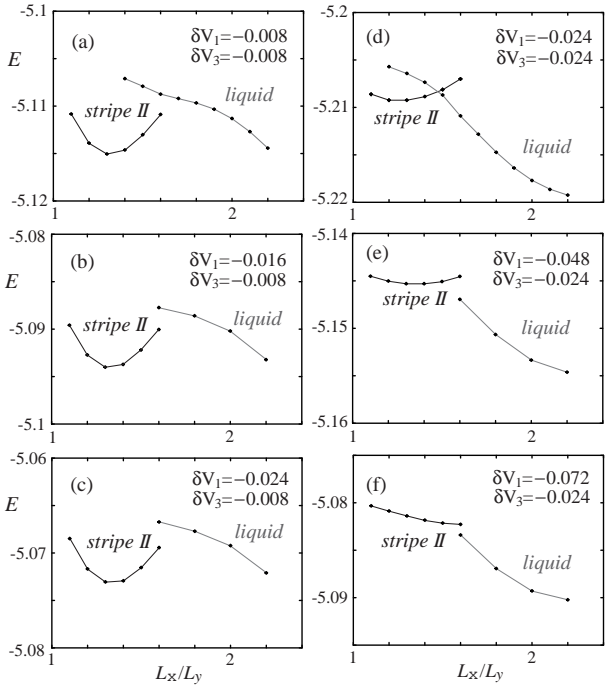}
\caption{The energies of the stripe state and the liquid state
for various Haldane's 
short-range pseudopotentials $V_1$ and $V_3$.
$\delta V_n$ is the difference from the value
of pure Coulomb interaction in the ideal two-dimensional system.
$\delta V_3=-0.008$ in (a)$-$(c), and it is $-0.024$ in (d)$-$(f). 
$N_e=12$. $\delta V_n=0$ for $n \ge 5$. }
\end{figure}

\section{Transition to a Liquid State}
Although the stripe state is expected in the 
numerical calculations, clear experimental 
evidence has not been obtained. 
In particular, transport properties of 
recent high-mobility wide quantum wells suggest
the liquid ground state around $\nu=1/4$,\cite{WQW}
while clear stripe correlations are obtained in the DMRG 
calculations. 
Since the previous DMRG study assumed the ideal 
two-dimensional system, we here consider the 
effect of finite width perpendicular to the 
two-dimensional layer.

When the wavefunction is enlarged in the
perpendicular direction, the short-range effective 
repulsion is reduced.
This reduction is represented by a decrease in 
Haldane's pseudopotentials $V_n$\cite{pseudo}, where 
$n$ is the relative angular-momentum between the
two electrons.
In the present study, we decrease the short-range components,
$V_1$ and $V_3$, and see how the ground state is modified.
In general, a decrease in $V_n$ for $n\ge 5$ 
is present, but the amount of the change rapidly decreases with 
the increase in $n$.
We therefore neglect the effects of $V_n$ for $n\ge 5$
in order to simplify the problem.

\begin{figure}[t]
\epsfxsize=80mm \epsffile{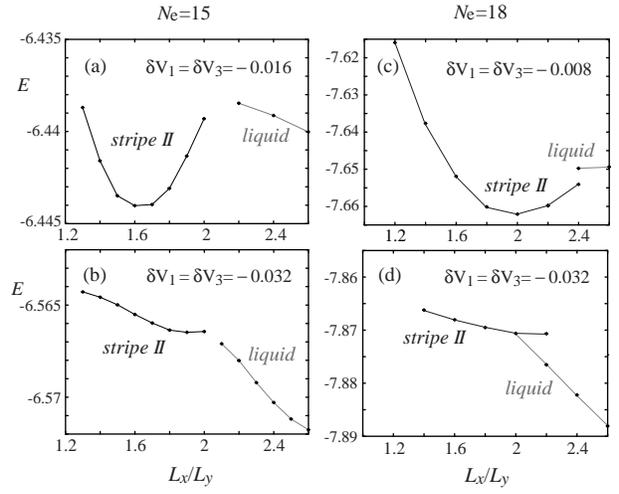}
\caption{The energies of the stripe state and the liquid state.
$\delta V_1$ and $\delta V_3$ are chosen to be the same.
$N_e=15$ in (a) and (b), and $N_e=18$ in (c) and (d).
}
\end{figure}

In Fig.~3, we plot the ground state energy for 
various $\delta V_1=V_1-(V_1)_{2D}$ and 
$\delta V_3=V_3-(V_3)_{2D}$ where 
$(V_n)_{2D}$ are the 
values for the pure Coulomb interaction in the ideal 
two-dimensional system. 
In this figure, we can see that $-\delta V_1$ 
increases the energy of both the stripe and liquid states. 
On the other hand, $-\delta V_3$ decreases the energy of the
liquid state compared with the energy of the stripe state; 
the stripe state at $\delta V_3=-0.008$ 
has lower energy compared with the liquid state 
around $L_x/L_y\sim 1.8$, while the liquid state apparently has 
lower energy for $\delta V_3=-0.024$. 
Furthermore, the aspect ratio dependence on the 
energy of the stripe state at $\delta V_3=-0.024$ is 
very small. 
These results indicate that the stripe structure at 
$\delta V_3=-0.024$ 
is not stable against changes in the period 
and the ground state becomes a liquid state which has lower 
energy. 
Thus we can expect a transition to 
a liquid state between $\delta V_3\sim -0.01$ and $-0.02$.
Since each state has different total momentum,
this transition is expected to be of the first order.

The existence of the transition is also found in larger
systems. 
Since the $V_1$ dependence on the
transition is small as shown in Fig.~3,
we consider the case of $\delta V_1=\delta V_3$.
The results for $N_e=15$ and 18 are
shown in Fig.~4, which reveals that
the liquid state at $\delta V_3=-0.032$  
has lower energy and the aspect ratio 
dependence on the energy of the stripe state is very small,
while the liquid state for $\delta V_3=-0.008$
has higher energy compared with the energy of the stripe state.
These features are the same as those of the system of $N_e=12$.
Since the size dependence on the transition
is small, we expect the transition to a liquid state
around $\delta V_3\sim -0.018$ even in the thermodynamic limit.

In order to see the effects on the ground-state wavefunction,
we calculate the ground-state pair correlation functions 
for $\delta V_1=\delta V_3=-0.008$, and $-0.032$.
The results are shown in Fig.~5,
where we can see that the stripe correlations 
shown in (a) of $\delta V_3=-0.008$ and (b)
of $\delta V_3=-0.032$
are almost the same, although the stripe state at 
$\delta V_3=-0.032$ has higher energy
compared with the energy of the liquid state as shown 
in Fig.~4 (d).
Since the correlation functions in the liquid state 
are completely different from those in the stripe state
as shown in Figs.~5 (b) and 5 (c),
the first-order transition between the stripe state and the liquid state
is confirmed.

\begin{figure}[t]
\epsfxsize=80mm \epsffile{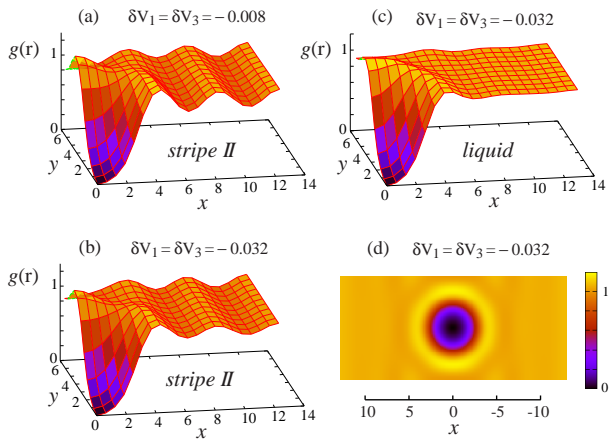}
\caption{
The pair correlation functions $g({\bf r})$ at
$\nu=3/8$ and $N_e=18$. (a)$-$(b) The type-II stripe state 
at $L_x/L_y=2.0$.
(c)$-$(d) The liquid state at  $L_x/L_y=2.2$.
}
\end{figure}

\section{Ground State Phase Diagram around $\nu=3/8$}
In order to investigate the phase boundary away from
$\nu=3/8$, we have calculated 
the ground-state energy and correlation functions at 
various fillings between $\nu=2/5$ and $\nu=1/3$ 
with various $V_n$. 
The obtained results are summarized in the 
phase diagram shown in Fig.~6. 
We find that the stripe state is stable around 
$\nu=3/8$ for small $|\delta V_n|$. 
The phase boundary to the liquid state 
moves toward smaller $|\delta V_n|$ 
as $\nu$ approaches the incompressible 
liquid states at $\nu=1/3$ and 2/5. 
The transition to the liquid state occurs even at 
$\delta V_n=0$ at $\nu\sim 0.35$ and $0.385$. 
These phase boundaries at $\delta V_n=0$ are
consistent with the previous DMRG calculations.\cite{Shib2}

\begin{figure}[t]
\epsfxsize=70mm \epsffile{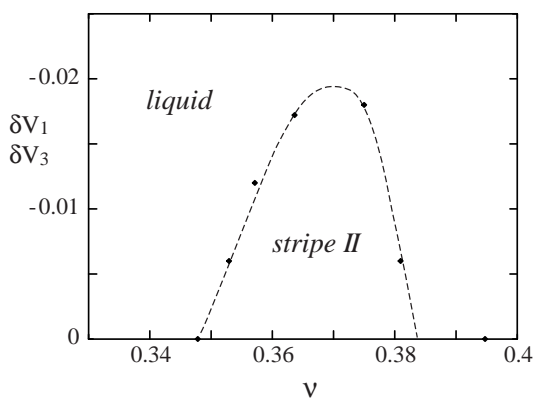}
\caption{The ground state phase diagram in the lowest Landau 
level between the incompressible
liquid states at $\nu=2/5$ and 1/3. $\delta V_1$ and $\delta V_3$
are chosen to be the same.
The broken line is a guide for the eyes.}
\end{figure}

Since the stripe state is not realized for large 
$-\delta V_1$ and $-\delta V_3$, the ground state
of wide quantum wells is expected to be a liquid state.
When we assume an infinitely deep quantum well and neglect
the effects from $\delta V_n$ for $n \ge 5$,
the critical width of about $6\ell$ is obtained.
In the case of realistic quantum wells,
the critical width may be smaller than this value
because the wavefunction penetrates through the potential
barrier.
The line of the phase boundary in Fig.~6 is just a guide for 
the eyes. If we precisely study the excitation spectrum 
around the boundary, some structures may be found.
This remains for future investigation.

\section{Summary}
In the present work we have studied the stripe 
state in the lowest Landau level using the DMRG method. 
The ground-state energy and the pair-correlation functions 
are systematically calculated 
at various fillings and pseudopotentials
around $\nu=3/8$. We have shown that the stripe state
in the lowest Landau level is realized in  
a system whose width perpendicular to the 
two-dimensional layer is sufficiently small.

\section*{Acknowledgement}
The present work is supported by
Grant-in-Aid No.~15740177 from MEXT Japan and No.~14540294 from JPSJ.


\begin{thebibliography}{99}
\bibitem{exp3}D.C. Tsui, H.L. St\"ormer and A.C. Gossard: 
Phys. Rev. Lett. $\bf 48$ (1982) 1559.
\bibitem{FQHE}R.R. Du, H.L. Stormer, D.C. Tsui, L.N. Pfeiffer and K.W. West:
Phys. Rev. Lett. $\bf 70$ (1993) 2944.
\bibitem{Lagh}R.B. Laughlin: Phys. Rev. Lett. $\bf 50$ (1983) 1395.
\bibitem{exp2}R.R. Du, H.L. Stormer, D.C. Tsui, A.S. Yeh, 
L.N. Pfeiffer and K.W. West: Phys. Rev. Lett. $\bf 73$ (1994) 3274.
\bibitem{jain}J.K. Jain: Phys. Rev. Lett. $\bf 63$ (1989) 199;
Phys. Rev. B $\bf 40$ (1989) 8079.
\bibitem{exp5_1}H.W. Jiang, H.L. Stormer, D.C. Tsui, L.N. Pfeiffer 
and K.W. West: Phys. Rev. B  $\bf 44$ (1991) 8107.
\bibitem{exp5_2}H.W. Jiang, R.L. Willett, H.L. Stormer, D.C. Tsui, 
L.N. Pfeiffer and K.W. West: Phys. Rev. Lett. $\bf 65$ (1990) 633.
\bibitem{exp5n}P.D. Ye, L.W. Engel, D.C. Tsui, R.M. Lewis,
L.N. Pfeiffer and K. West: Phys. Rev. Lett. $\bf 89$ (2002) 176802.

\bibitem{yoshi}D. Yoshioka: Phys. Rev. B $\bf 29$ (1984) 6833.
\bibitem{Halp}B.I. Halperin, P.A. Lee and N. Read:
Phys. Rev. B $\bf 47$ (1993) 7312.
\bibitem{Onod}M. Onoda, T. Mizusaki, T. Otsuka and H. Aoki:
Phys. Rev. Lett. $\bf 84$ (2000) 3942.
\bibitem{Lam}P.K. Lam and S.M. Girvin: 
Phys. Rev. B $\bf 30$ (1984) 473.
\bibitem{Kun}Kun Yang, F.D.M. Haldane and E.H. Rezayi:
Phys. Rev. B $\bf 64$ (2001) 081301.

\bibitem{Shib2}N. Shibata and D. Yoshioka: 
J. Phys. Soc. Jpn. $\bf 72$ (2003) 664.

\bibitem{CFSt}
S.-Y. Lee, V.W. Scarola and J.K. Jain: 
Phys. Rev. Lett. $\bf 87$ (2001) 256803.

\bibitem{WQW}
W. Pan, H.L. Stormer, D.C. Tsui, L.N. Pfeiffer, K.W. Baldwin 
and K.W. West: Phys. Rev. Lett. $\bf 88$ (2002) 176802.

\bibitem{DMRG}S.R. White: Phys. Rev. Lett. $\bf 69$ (1992) 2863;
Phys. Rev. B $\bf 48$ (1993) 10345.
\bibitem{Shib}N. Shibata and D. Yoshioka: 
Phys. Rev. Lett. $\bf 86$ (2001) 5755.
\bibitem{DY}D. Yoshioka and N. Shibata:
Physica E $\bf 12$ (2002) 43.
\bibitem{Shib3}N. Shibata: 
J. Phys. A: Math. Gen. $\bf 36$ (2003) R381.
\bibitem{pseudo} F.D.M. Haldane and E.H. Rezayi:
Phys. Rev. Lett. $\bf 54$ (1985) 237.

\end{thebibliography}
\end{document}